\documentclass[preprint,12pt,3p]{elsarticle}

\usepackage{amssymb}
\usepackage{color,soul}
\usepackage{graphicx}
\usepackage{float}
\usepackage{enumitem}

\newcommand{\chiles}{{\sc chiles}}
\newcommand{\split}{{\sc split}}
\newcommand{\clean}{{\sc invert}}

\newcommand{\strace}{{\sc strace}}
\newcommand{\proc}{{\sc proc}}
\newcommand{\casapy}{{CasaPy}}

\newcommand{\mset}{{MeasurementSet}}
\newcommand{\msets}{{MeasurementSets}}
\newcommand{\ple}{{\em Pleiades}}
\newcommand{\mgs}{{\em Magnus}}
\newcommand{\aws}{{\em AWS}}

\journal{Astronomy and Computing}

\begin{document}

\begin{frontmatter}

\title{Imaging SKA-Scale data in three different computing environments 
}

\author[icrar]{Richard Dodson\corref{cor1}}
\address[icrar]{International Centre for Radio Astronomy Research (ICRAR),\\
The University of Western Australia,\\
M468, 35 Stirling Highway, Crawley, Perth, WA 6009, Australia }
\cortext[cor1]{Corresponding Author}
\ead{richard.dodson@icrar.org}

\author[icrar]{Kevin Vinsen}
\ead{kevin.vinsen@icrar.org}

\author[icrar]{Chen Wu}
\ead{chen.wu@icrar.org}

\author[caastro]{Attila Popping}
\ead{attila.popping@icrar.org}

\author[icrar]{Martin Meyer}
\ead{martin.meyer@icrar.org}

\author[icrar]{Andreas Wicenec}
\ead{andreas.wicenec@icrar.org}

\author[icrar]{Peter Quinn}
\ead{peter.quinn@icrar.org}

\address[caastro] {Australian Research Council, Centre of Excellence for All-sky Astrophysics (CAASTRO)}

\address[columbia]{Department of Astronomy, \\
Columbia University,\\
Mail Code 5246, 550 West 120th Street,\\
New York, New York 10027, USA}
\author[columbia]{Jacqueline van Gorkom}
\ead{jvangork@astro.columbia.edu}

\address[nrao]{National Radio Astronomy Observatory, \\
1003 Lopezville Rd., P. O. Box O, \\
Socorro, NM 87801, USA}
\author[nrao]{Emmanuel Momjian}
\ead{emomjian@nrao.edu}

\begin{abstract}

We present the results of our investigations into options for the computing platform for the imaging pipeline in the \chiles\ project, an ultra-deep HI pathfinder for the era of the Square Kilometre Array. \chiles\ pushes the current computing infrastructure to its limits and understanding how to deliver the images from this project is clarifying the Science Data Processing requirements for the SKA. 
We have tested three platforms: a moderately sized cluster, a massive High Performance Computing (HPC) system, and the Amazon Web Services (AWS) cloud computing platform. 
We have used well-established tools for data reduction and performance measurement to investigate the behaviour of these platforms for the complicated access patterns of real-life Radio Astronomy data reduction.
All of these platforms have strengths and weaknesses and the system tools allow us to identify and evaluate them in a quantitative manner. 
With the insights from these tests we are able to complete the imaging pipeline processing on both the HPC platform and also on the cloud computing platform, which paves the way for meeting big data challenges in the era of SKA in the field of Radio Astronomy.
We discuss the implications that all similar projects will have to consider, in both performance and costs, to make recommendations for the planning of Radio Astronomy imaging workflows.


\end{abstract}


\end{frontmatter}

\section{Introduction: Developments towards data-driven Radio Interferometric processing}

In this paper we present part of our on-going efforts towards prototyping `Data-Driven Processing' for the Square Kilometre Array (SKA) Science Data Processor (SDP).
The SKA will require advances of several orders of magnitude in the processing of data, pushing Radio Astronomy into the forefront of the `Big-Data challenge' \citep{pq_ska}. 
The SKA Phase 1 data-rates out of the correlator will approach a Terabyte/Sec 
for SKA-MID and SKA-LOW \citep{ska1_rebase} combined. 
The final stage, with the complete collecting area, will be an order of magnitude higher. 
The calculation of the required data processing rates for the SDP is highly dependent on the science case, but will be several hundred PetaFLOPS in the most extreme cases. 
To deliver such performance, highly parallel computer processing solutions are required and in this paper we set out to explore some of the options. 

In this investigation we are testing pipeline solutions for the calibrated and flagged datasets from the Karl G. Jansky Very Large Array (VLA) deep HI survey \chiles, the COSMOS HI Large Extragalactic Survey \citep{fernandez2013pilot,chiles}.
{This survey aims to study the neutral atomic hydrogen (HI) content of galaxies over 4 billion years of cosmic time, approximately 1/3 the history of the Universe and twice the lookback time of any previous emission-line survey.  HI is a crucial ingredient to study for understanding galaxy evolution as it is the dominant baryonic fuel out of which stars and galaxies are ultimately made, as well as being an important tracer of galaxy kinematics.  Such surveys have previously been too expensive to carry out due to limitations in telescope technology and back-end processing resources.}
The \chiles\ survey will be one of the prime pathfinders for the data processing on SKA scales. 
{(Although not alone in this; the LOFAR project \citep{van2013lofar} also has similar challenges.)}
The data volumes and processing requirements mean this project will stretch the bounds of current computing capability.
The \chiles\ project therefore is performing a crucial role in our prototyping investigations for the key SDP concepts and approaches. 

\subsection{The \chiles\ project}

\chiles\ is running at the VLA, which is a 27 antenna array. 
The new, upgraded front-end (wide-band L-band receivers) and back-end (the WIDAR correlator) \citep{evlamemo_152,evlamemo_165}, provided through the Expanded Very Large Array project \citep{vla_backend}, can now provide instantaneous coverage for spectral line observing between $\sim$940 and $\sim$1430-MHz on the sky (15 spectral windows of 32MHz, each giving a total of 480MHz in each session). 
The observations are dithered in frequency to smooth out the edges of the spectral windows.
The antennas (being 25m in diameter) see about 0.5 degrees ($\sim$2000 arcsecond) across the sky at the pointing centre. 
The array configuration is VLA-B, which has 11\,km baselines and a typical beam size of $\sim$5x7 arcsec at 1.4 GHz, assuming natural weighting. We are currently oversampling this with 2048 pixels of 1 arcsecond in size during this development phase. 
This data is in 15.625kHz channels (representing about 3km/s at the rest frequency of the HI line being observed). 
Therefore there are 351 baselines, a little less than 31,000 channels per polarization product to be processed, and a full field of view of  2048x2048 pixels in the image plane. 

These datasets are therefore much larger than those normally analysed and therein lies the challenge. 
We expect about five epochs of observing (the epochs are defined by when the VLA is in the correct configuration for the science, approximately every 15 months).
The first epoch was of 178 hours in total, broken into 42 days worth of observing, with each day's observation between 1 and 6 hours long.
The mean size of the flagged and calibrated dataset from a days observations in the first epoch are 330GB, with the maximum size being 803GB and the minimum being 45GB.
The second epoch of 213 hours has been observed but not yet calibrated. 
One of the main challenges for the project will be to produce the image cubes from the observations, post-flagging and calibrating.
The processing steps required to complete this analysis will be presented in a forthcoming paper; here we limit ourselves to the investigations on the computing resources required to undertake the analysis.

\section{Issues for SKA-Scale datasets}

\subsection{Compute environments}

To investigate the application and total costs of operation for the workflow in a range of indicative environments we have repeated the same data-reduction pipeline on three very different computing infrastructures: a moderate sized cluster (\ple), such as a group like ICRAR could (and does) host and control; a high performance computing cluster (\mgs) that would be provided by a national facility 
such as the  Pawsey centre \citep{website:pawsey} and a cloud computing environment, such as provided by the Amazon Web Service (\aws). 
This allowed us to explore three very different approaches, all of which would be of the scale accessible to groups such as ours via in-house capital expenditure, via competitive applications for resources on national infrastructure or via cumulative operational expenditure, respectively. 
{It should be noted that the SKA infrastructure is not necessarily limited to the above three candidate environments, so these may not represent the final choice of SKA architecture.}

\subsection{Data Transfer}

We have found that the copying stage is an important work item, which is not normally considered part of the data reduction pipeline. 
Given the size of the input and output data items, we tried to keep the data movement to a minimum.

The data was moved from the CHILES project's data repository at NRAO (Socorro) to Perth for every individual observation, after the flagging and calibration steps had been completed.
Once the data was in Perth it was stored on a large dedicated storage pod and then copied to the target test environments of \aws\ and \mgs.
The storage pod is connected via a 10G ethernet connector.

The copying stage, as it is immediately followed with the splitting, could be combined into a single process or be staged via temporary short term storage.

\subsection{Data reduction tools}
\label{sec:drt}

The data-reduction tools we use are exclusively from \casapy\ \citep{mcmullin2007casa} Version 4.3 and therefore the issues that arose
all revolve around limitations in the CASA performance, as discussed in the following sections. Note that `operations' are printed in small capitals; `tasks' used to perform the operations are printed in small capitals with brackets appended.

{The data provided by NRAO has been calibrated and the target field selected, therefore the remaining operations to be performed in the HI-data reduction are: to copy the data to an accessible point, to split the data into manageable sizes whilst correcting for the station doppler shifts of that day and selecting specific frequency ranges and to Fourier invert the observed data (taken in the reciprocal of the image domain) into a 3-D image cube (these dimensions being Right Ascension, Declination and Velocity). Additionally one should deconvolve the image to correct the initial `dirty' image for the spatially extended point spread function (PSF), which is the Fourier Transform of the points in reciprocal space where the antenna pairs measured the correlated signal. After the deconvolution the PSF is replaced with a compact Gaussian representing the maximum resolution to form the `clean' image. See \citet{tms} for a full discussion of these concepts. 
Traditionally one would read all the data files simultaneously and invert to produce an image cube, but this is not possible as the task {\sc clean()} \citep{Hogbom:1974uf} fails due to the extreme size of the input datasets.
This is why we must split the input data into smaller frequency ranges, or sub-bands, and perform \clean\ on all of the days simultaneously, but with fewer channels. }

The copying does not involve data reduction, so is discussed in detail later.
Once the data is accessible we need to pre-select the data from the input \msets (henceforth referred to as the operation \split). 
For this we have trialed the tasks {\sc split()}, {\sc cvel()} and finally {\sc mstransform()} and have settled on the latter, as that was the fastest. %
We note that the doppler shift correction process involves FFTing the entire selected frequency range (or spectral window) before applying the doppler shift, then selecting only those channels which fall in the requested frequency range. 
This process has a strong potential for improved efficiency. 

The \clean\ operation takes the frequency split data and combines the many days into images for that frequency range. In these investigations we have not deconvolved the images, except to investigate the residual noise as discussed in Section \ref{sec:resid}.

We found that with the limited frequency ranges we were using, the noise levels per channel were very sensitive to the weighting scheme. For example uniform weighting (or Briggs weight -2) causes large increases of noise at the edges of the sub-cube. This is due to the implementation of the Briggs' scheme in \casapy, which depends on the total data selected not just the data channel being imaged. Obviously in this case, where we have a limited input frequency span of data points from which to derive the weights, the edge data displays enhanced noise. 
However with natural weighting (equivalent to Briggs +2), or any Briggs weight greater than $\sim$1 this does not occur. Alternative weighting schemes are being discussed, both in the \casapy\ team and within the literature (see \citet{boone_13, yatawatta_14}) so this should not be an issue for the SDP continuum imaging pipeline architectures. 


The final operation is to combine the individual small image cubes into the required full sized cube. 
The final cube is 2048$^2$ pixels by 30,720 channels, single stokes, resulting in a $\sim$500GB cube when using 4 bytes per voxel. 

In these tests we have not explored the range of options for deconvolution and continuum subtraction, as the detailed plan is still being developed. 
This latter step could be post-clean (distributed but frequency independent) or post-combination. 
In the latter case the parallelisation could come from subtraction along individual pixels of the signal averaged (perhaps with a spectral dependence) along frequency.

\subsection{Parallelism}
The obvious issue for massive datasets is how to process them in a parallel fashion. 
It is an often-stated fact that ``Astronomy data is embarrassingly parallel''; image cubes can be formed per frequency channel and calibration can be performed per solution interval. These are to a large extent completely independent of each other. 
However when working with the standard data formats, be they FITS or \msets, the combination of the visibility data into a single file or file structure limits the immediate implementation of the natural parallelism. 
Therefore we have derived a workflow, see Figure \ref{fig:workflow_image2},  which would divide the data between work-units in a usable fashion. 
The input data was per day, and this was used as the division in the first stages ({\sc copy} and \split). 
After splitting the day's data into multiple frequency sub-bands the images can be formed in parallel by combining all the days across the different frequency sub-bands simultaneously. 


\section{Methods}

In this section, we discuss the workflow for the analysis pipeline and how we assessed it. 

\subsection{The proposed data flow}

\begin{figure}[hbt]
    \centering
    \includegraphics[width=0.9\textwidth]{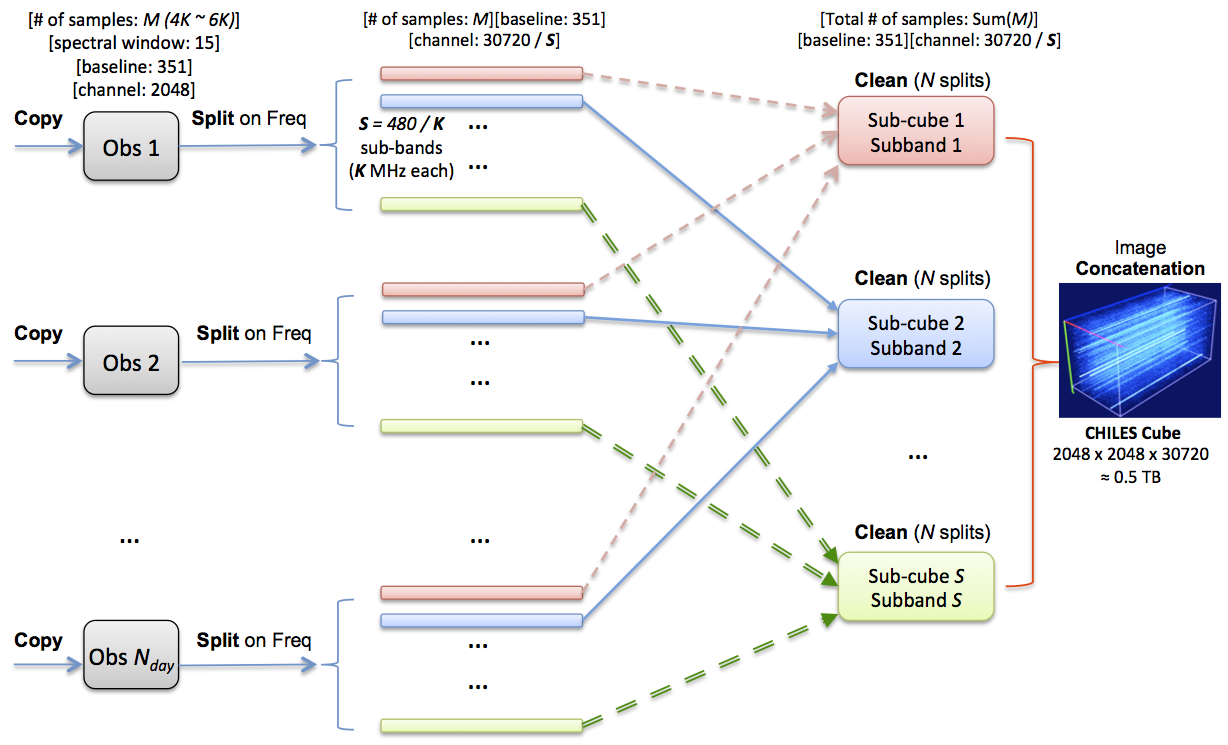}
    \caption{A data distribution view of this workflow. This also shows all information on data dimensions and indices that are important for our scheduling data partitioning, distribution and gathering operations. \(M\) represents the number of visibilities collected per time step, ranging from $\sim$4,000 to $\sim$6,000. The number of sub-bands \(S\) is determined by the instantaneous bandwidth (i.e. 480 MHz) and each subband's frequency width \(K\) (e.g. 4MHz).}
    \label{fig:workflow_image2}
\end{figure}

The detailed workflow model we are following is shown in
Figure \ref{fig:workflow_image2}, which represents the data distribution breakdown.
For each day (that is N$_{\rm day}$ parallel processes) there is a copy process (or the ingest stage), followed by a split process to separate each frequency. 
The ingested \msets\ could be retained in a temporary archive. 
The split includes the data re-ordering, as the outputs are files that are frequency sorted as well as divided by day. 
The \clean\ process (that is N$_{\rm split}$ parallel processes) takes the frequency split data and combines the many days into images for that frequency range. 
The final process is the combination of the frequency ranges into a concatenated cube.


\subsection{The script layout}

The work flow is controlled by a set of python and shell scripts.
The shell scripts contain the setup information (and are provided to the queue) and the python scripts extract that setup information to drive the process.
The pipeline scripts for the three environments are extremely similar, but have to be independent because of the different processing environments.

\subsection{Results Capture}

We use two different methods to collect performance metrics for both compute (e.g. CPU and memory usage) and I/O (e.g. I/O operations, throughput, inter-arrival time, etc.). 
In the first method, we periodically measure a list of process-specific kernel counters available in the Linux \proc\ file system \citep{Kernel-counter_09} while the processing tasks were running. 
The sample interval is currently one measurement per second. 
While this method provides useful measurements on CPU and memory usage, some detailed I/O metrics cannot be directly derived from the \proc\ file system. Therefore, we used a second method --- the \strace\ \citep{Strace_01} linux system tool --- to access more advanced I/O performance indicators, such as whether the I/O requests were sequential or random and the size of each I/O requests issued to the underlying file system. \strace\ is able to capture all system calls and signals. However, since we are only interested in I/O requests made by the application as system calls, we instruct \strace\ to only measure four types of system calls --- file descriptor related, process management-related (in order to track sub-processes), socket-related and those which take a file name as an argument such as open, close, read, write, etc. One issue of \strace\ is the tracing overhead associated with frequent context switching, which can vary between less than 10\% (for hundreds of system calls per second) and over 100\% (for tens of thousands of system calls per second). This overhead in turn substantially prolongs the application completion time. However, this is not an issue for profiling the I/O access patterns (random and sequential), which are basically time invariant. The \proc\ and \strace\ measurements are both provided in the additional data products available in the online version of this paper and only the results from \proc\ are plotted in the printed version.

\section{Test Environments}

Three test environments were selected to represent three different approaches to the data reduction. 
Our goal is to present options for deciding which model for sourcing computational resources would best match both the specific case we address (i.e. \chiles) and guide the resourcing decisions that will be raised for other computing problems.

\subsection{Moderate Size Departmental Cluster Platform}


\ple\ was specifically designed and built to provide a development platform for ICRAR's HPC projects. It is a six-compute node HPC cluster (+1 head node) located at ICRAR. 
Each of the compute nodes currently contains a dual Intel Xeon X5650 2.66GHz CPU, 64-192 GB of RAM, one Tesla or two K10 or K20 GPUs, and a  Mellanox MT26428 QDR (40Gbps) Infiniband interconnect. 
%
In addition, a dedicated storage node provides persistent data across the QDR Infiniband fabric. 
The \chiles\ dataset is provided on a triple RAID-6 striped volume that is 147TB in size. 

\subsection{High Performance Computing Platform}
\label{label:hpc}
The \mgs\ HPC cluster is provided by the Pawsey Centre, which plays a key role in the Australian Government’s strategy to provide high level scientific computing resources for the Australian research community. 
It is sited in Perth, owned by CSIRO and managed by the Pawsey Supercomputing Centre. 

\mgs\ comprises 1536 nodes in 384 blades. Each compute node hosts two 12-core, Intel Xeon E5-2690V3 “Haswell” processors running at 2.6 GHz, for a total of 35,712 cores, delivering in excess of 1 PetaFLOP of computing power. Each node hosts 64GB of RAM.
The nodes communicate amongst themselves over Cray's high-speed, low-latency Aries interconnect. 
Global storage (also known as the scratch file system) is provided by a three-cabinet Cray Sonexion 1600 Lustre appliance, with a usable capacity of 3PB and a sustained read/write performance of 70 GB/sec.
In the November 2014 Top500 list, \mgs\ debuted at \#41, achieving 1,097 TeraFLOPS (1 PetaFLOP+). 
At the time of writing, this makes \mgs\ the most advanced scientific supercomputer in the Southern Hemisphere.
For our investigations we requested 44-nodes of \mgs , 
{about 3\% of the total computing power. This ratio is carried forward for the calculation of the fractional capital expenditure.}

\subsection{Cloud Computing Platform}
The cloud environment allows for considerable flexibility, which is discussed below;
The main constraint is how much one is willing to pay for the performance. 
The code to run the Pipeline on \aws\ is written in Python using the boto package~\cite{website:boto}. 
This allows us to start many servers with different configurations on demand when we need them. 
Our python scripts will always look for the cheapest spot price in the regions specified.

\subsubsection{Disk Storage}
Disk storage is provided by Amazon Elastic Block Store (EBS).
Amazon EBS volumes are network-attached, and persist independently from the life of an Amazon Elastic Compute Cloud (EC2) instance.  
Amazon provides three volume types: General Purpose (SSD), Provisioned IOPS\footnote{Input/Output Operations Per Second} (SSD), and Magnetic. 
General Purpose (SSD) is the SSD-backed general purpose EBS volume type. 
IO rates are primarily controlled by the instance types generic network capacity. 
Provisioned IOPS (SSD) volumes offer storage with consistent and low-latency performance. 
These were used for all EBS instances to improve the IOPS required by \casapy. 
Initial experiments showed the general purpose SSD gave about 20-30 IOPS with \casapy. 
Provisioned IOPS improved this to between 90-100 IOPS.
Magnetic storage was not used.

Many Amazon EC2 instance types can also access disk storage located on SSD disks that are physically attached to the host computer and do not persist. This disk storage is referred to as instance store or ephemeral storage. 
This was used for scratch storage for some processing tasks and when data was to be written to long term storage in the Amazon Simple Storage Service (S3).

\subsubsection{Long Term Storage}
Amazon S3 provides access to a reliable data storage infrastructure. 
S3 stores data as objects within resources called ``buckets''. 
One can store as many objects as required within a bucket, and write, read, and delete objects in the bucket. 
Objects can be up to 5TB in size.

S3 is designed for 99.999999999\% durability and 99.99\% availability of objects over a given year. 
There is also a low-cost Reduced Redundancy Storage option for less critical data, and Amazon Glacier for long term storage where access time is not important. 
All our work used the reduced redundancy S3 storage.
\subsubsection{Spot instance vs On-Demand Instance pricing}
AWS has two relevant pricing models. 
A third option exists call Reserved Instances, but that requires the purchase of 1 or 3 year contracts and was not used for these tests.
\begin{description}
\item[On-Demand Instances:] These provide the purchase of compute capacity by the hour with no long-term commitments or upfront payments. 
One can increase or decrease the compute capacity depending on the demands of the application and only pay the specified hourly rate for the instances  used. 
\item[Spot Instances:] These provide the ability to purchase compute capacity at hourly rates, usually at lower cost than the On-Demand rate. 
Spot Instances allow us to specify the maximum hourly price that we are willing to pay to run a particular instance type. 
EC2 sets a Spot Price for each instance type in each Availability Zone, which is the price all customers will pay to run a Spot Instance for that given period. 
The Spot Price fluctuates based on supply and demand for instances, but customers will never pay more than the maximum price they have specified. 
If the Spot Price moves higher than a customer’s maximum price, the customer’s instance will be shut down after a two minute warning. 
Table \ref{tab:price_comparison} shows the difference in cost between on demand and spot prices at the AWS Sydney data centre during the test runs. For the final processing only spot instances, being significantly cheaper, were used.
\end{description}

\begin{table}[hbt]
\centering
\begin{tabular}{l | c c}
Instance & On demand (AUD) & Spot Price (AUD) \\
\hline
\hline
m3.medium & \$0.098 & \$0.01 \\
m3.xlarge & \$0.392 & \$0.04 \\
r3.2xlarge & \$0.840 & \$0.09 \\
r3.4xlarge & \$1.680 & \$0.20 \\
\end{tabular}
\caption{A table showing the typical difference in cost between on demand and spot prices on the AWS cloud. These numbers are for the Sydney data centre on 6 Mar 2015}
\label{tab:price_comparison}
\end{table}

\section{Operational Flow}

\subsection{Cluster and HPC Environments}

Both the cluster and HPC environments were sufficiently similar that they can be described as a common work flow. 
The hardware resources required for each step of the work flow are requested via PBS or SLURM (for \ple\ or \mgs\ respectively). 
The work is then divided between the requested cores. 
This is one of the major issues for the workflow with a conventional cluster, as there is no flexibility in the differing hardware requirements during the workflow. 
The parallel resources can either be set to match the number of days, or the number of frequency sub-bands. 
If either is smaller than that required, the work-units will cycle over all those stages until all the tasks are completed. 
Whilst waiting for a particularly slow work-unit or for a work-unit which has been allocated more work than the others the requested resources are held without being active. 
We can sub-divide the stages for more efficient a-priori allocation of resources, but this still falls foul of work-units which take longer than the average. 

\subsubsection{Step 1 - Copy}

For the cluster/HPC environments we had all the initial data products in place on globally accessible storage. 
Given that the first epoch (including the pre-calibrated data) is around 80TB, and future epochs will be larger, this is a significant storage requirement. 
The cluster/HPC environments are a fixed hardware configuration and the configuration of the required software (mainly the \casapy\ setup and the cloning of the pipeline software from github) is straight forward. 
As described in section \ref{label:hpc}, \mgs\ has a luster file system, however \ple\ has stored the data on an external raid network attached disk. This limited filesystem was a significant roadblock for the parallel implementation of the data distribution in this environment.  

\subsubsection{Step 2 - Splitting the files}

The list of days to be processed is counted and these are divided between the work-units. Where the number of days is greater than the number of work-units, each work-unit is allocated multiple days to process. 
Each work-unit is responsible for splitting one \mset, as \casapy\ by default locks a dataset from further access. The work-unit cycles over the requested frequency range, breaking the data into the smaller more manageable sub-bands. These sub-bands are stored for the next stage. 
As \casapy\ does not support concurrent access to large \msets\ it would be possible to significantly improve the performance of this step, for example with a parallel splitting implementation of {\sc mstransform()}.

\subsubsection{Step 3 - Inverting the image}

The \clean\ stage combined all the days for a particular frequency sub-band  into  a single image and (if requested), deconvolves this image with a conventional {\sc clean()} \citep{Hogbom:1974uf} step. 
Until the work flow is finalised we have only performed a limited deconvolution of the datasets; currently we only clean the image in each individual frequency channel (i.e. 15.625\,kHz) rather than attempting to subtract continuum sources over the entire frequency range. 

\subsubsection{Step 4 - Image Concatenation}

All the images of the frequency sub-bands are recombined into a larger image cube. We have found that \casapy\ is unable to combine the entire frequency range for the entire Field of View.
We  add together only 12 sub-bands as the resulting \mset\ is 48GB; anything larger becomes too large to open in the standard visualisation tools and to combine all the datasets causes \casapy\ to crash.

\subsection{EC2 Configuration}
Table \ref{tab:ec2_configuration} shows the 4 different types of EC2 instance that were used in the workflow.
The main drivers for selecting a particular instance type where the memory required to run \casapy\ and the ephemeral storage required to hold the intermediate data.
For example {\sc clean()} cannot use 16 cores, but we required the 320GB of fast, direct attached SSD to make the process run quickly.

\begin{table}[hbt]
\centering
\begin{tabular}{l l | c c c}
Model & Used on & vCPU & Memory (GB) & Instance Storage (GB) \\
\hline
\hline
m3.medium & copying & 1 & 3.5 & 1 x 4 \\
m3.xlarge & splitting & 4 & 15 & 2 x 40 \\
r3.2xlarge & concatenation & 8 & 61 & 1 x 160 \\
r3.4xlarge & cleaning & 16 & 122 & 1 x 320 \\
\end{tabular}
\caption{The \aws\ instance types used for various stages of the work flow: Copying, Splitting, Cleaning and Concatination. A vCPU is equivalent to a single CPU thread. Instance Storage is the SSD ephemeral disk storage.}
\label{tab:ec2_configuration}
\end{table}

\subsubsection{Step 1 - Data transfer - Initial Setup}
\begin{figure}[h!bt]
    \centering
    \includegraphics[width=0.8\textwidth]{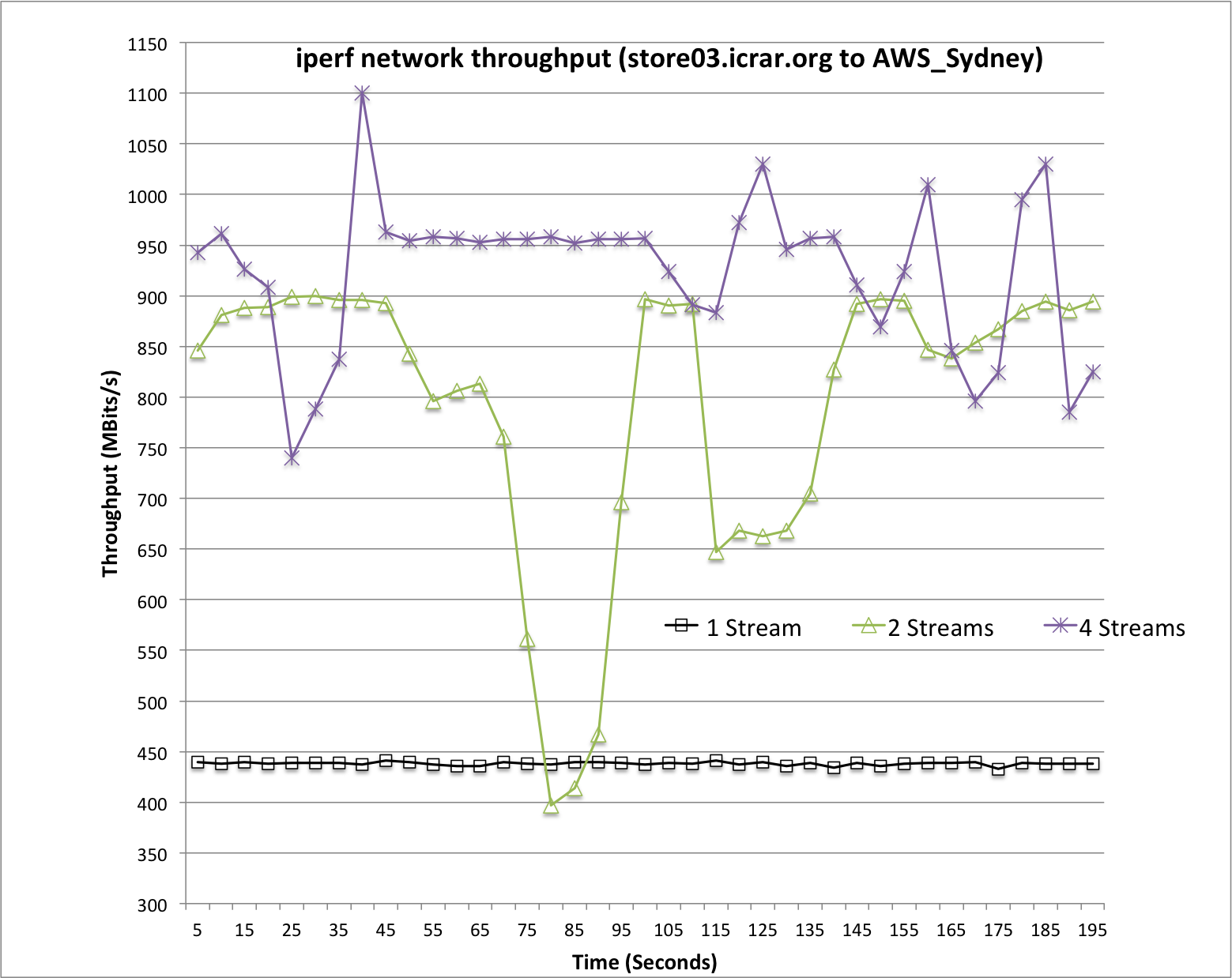}
    \caption{iPerf network throughput tests for 1, 2 and 4 parallel streams. It shows that 2 or more parallel streams saturate the connection to \aws , suggesting the limit of the bandwith is 1Gb. Therefore for the data transfer we limited ourselves to 2 streams.}
    \label{fig:aws_network_thruput_img}
\end{figure}

To transfer the 10+TB of calibrated visibility datasets, over WAN to AWS Sydney data centre in an efficient manner, we created a dedicated ``copy machine" (using an m3.medium instance). 
We then fine tuned the Linux kernel configuration on both ends of the link based on \citet{web:kernel_tune-site}. 
In particular, we increased the \textit{net.core.rmem-max} value to 1600 MB in order to accommodate the large bandwidth delay over the WAN link from ICRAR to AWS Sydney. 
The network iperf tests (Figure \ref{fig:aws_network_thruput_img}) show that 2-4 parallel streams will saturate the ICRAR-AWS network bandwidth, which appears to be 1Gb. 
We therefore employed the {\sc bbcp} \citep{web:BBCP-site} transfer tool to send two parallel data streams for all the subsequent data transfer.
EBS snapshots of each day's observations were created by copying the \msets\ from the data storage area in ICRAR onto an EBS volume, whose size had been pre-determined to accommodate the \mset\ files.
Once the data transfer was complete the EBS volume was detached from the copy machine, the snapshot was created and the EBS volume was deleted.
As \aws\ bill for the EBS volumes per GB, once a volume was no longer required it was deleted.

We chose to create EBS snapshots rather than placing the data in the S3 archive because creating a new EBS volume from a snapshot takes $\sim$3 seconds, so we were able to create multiple parallel instances and attach different volumes to each, to split the 480\,MHz \mset\ into smaller \msets\ spanning 4MHz. 
We could not mount the \mset\ on a shared disk and have parallel access to it because \casapy\ does not support concurrent access to large \msets.

\subsubsection{Step 2 - Splitting the files}
As we are paying for the spot instance by the hour (or part thereof) it is more cost efficient to split the file into 4-5 separate \msets\ one after the other. 
The system creates an EC2 instance in the spot market and then uses cloud-init and bash scripts to attach a new EBS volume created from the appropriate snapshot. 
Each instance then splits the days \mset\ into the smaller frequency sub-bands. 
The degree of parallelism is controlled by the user and is driven by how quickly the results are required.
Anything under 3 sub-bands is not particularly economic as 100GB measurements sets can be done in under an hour. 
The 700GB \msets\ could take an hour to process. 
The resulting \msets\ are then copied to S3 for loner term storage, as that provided the best balance of cost and accessibility for our requirements.

\subsubsection{Step 3 - Inverting the image}

\begin{figure}[hbt]
    \centering
    \includegraphics[width=0.9\textwidth]{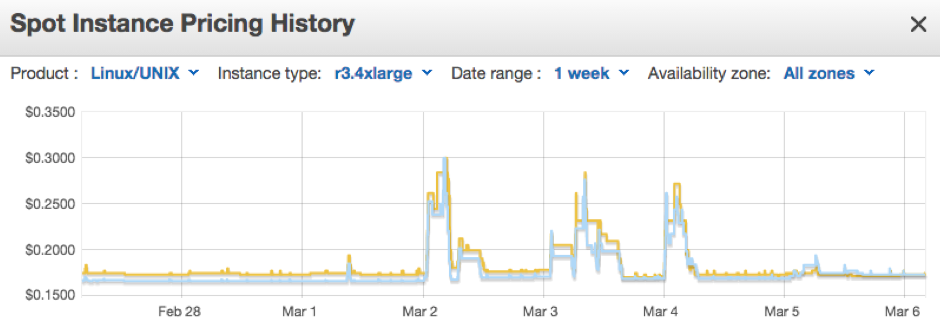}
    \caption{The variations in Spot Price for March 2015 in the \aws\ Sydney as the multiple instances of \clean\ were launched, as provided by the \aws\ monitoring tools. The baseline price (y-axis) for a r3.4xlarge instance was about \$0.16, but when we submitted multiple parallel tasks the price was driven almost double. We therefore were careful to submit the clean jobs in a staggered fashion, manually}
    \label{fig:spotprice}
\end{figure}

The EC2 \clean\ operation downloads the 4MHz frequency \msets\ for each of the daily datasets from the S3 archive and stores it on a large ephemeral disk for local processing. 
After all the days have been combined into one image cube spanning the frequency sub-band, it was written back to S3.
Because the \clean\ requires a big \aws\ instance it affects the spot market price quite quickly; 
therefore 
%
{for these tests the scheduling of the runs on \aws\ was done by hand.
The spot price was checked using the \aws\ dashboard and the parallel tasks launched one by one.
Running the processes strongly affected the spot price in the Sydney \aws\ region as can be seen by Figure~\ref{fig:spotprice}.
This meant we had to manually throttle the starting of tests to let the price return to an acceptable level.}

\subsubsection{Step 4 - Image Concatenation}

The cleaned \msets\ are copied back from S3 and concatenated, in 48MHz sections for the reasons above. Once these are formed we archive the final data products on S3 from where they can be downloaded to the host, or any other collaborating, institution.

\section{Results and Discussion} 



The pipelines were deployed on all three platforms, both as single broken-out work-units and as parallel work-units. The single work-units functioned as expected on all platforms and provided the following measurements and analysis of the system performance metrics. It is important to stress that the parallel processing did not proceed on \ple, as we immediately hit the network-attached disk access limits. The complete process can be performed on \ple, but only in a serial fashion. Therefore this platform must be considered to be unable to deliver the required workflow. This is of course compatible with the role of this cluster, which is only designed for workflow prototyping before implementation on larger HPC platforms. On \mgs\ the total completion time was $\sim$110 hours of run time, after $\sim$170 hours in the queue. On \aws\ the total completion time was $\sim$96 hours of run time, manually staggered to avoid driving up the spot market.

\begin{figure}[h!bt]
    \centering
    \includegraphics[width=1.0\textwidth]{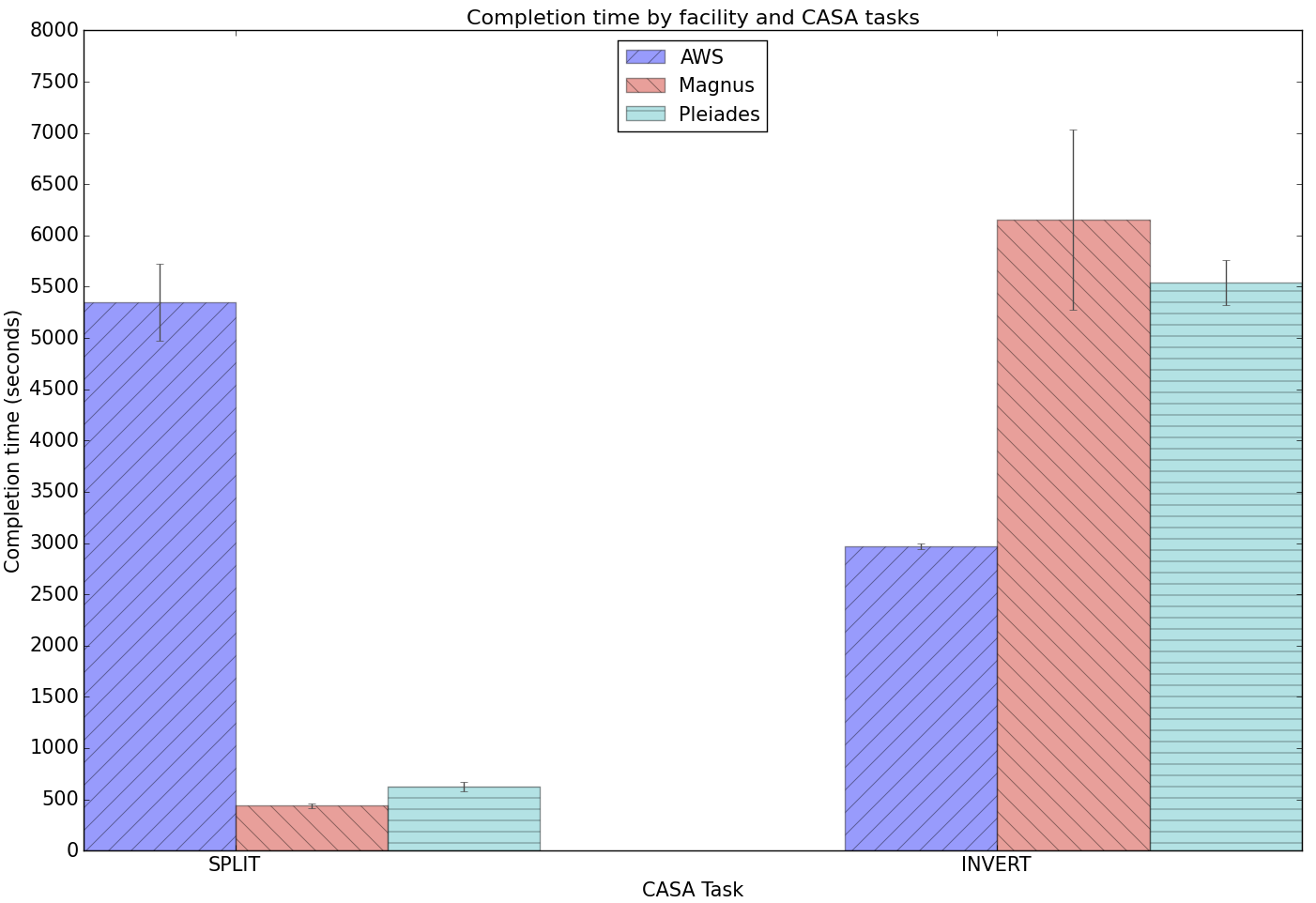}
    \caption{Comparison of three facilities on the two main CASA tasks: \split\ (i.e. {\sc mstranform()}) and \clean\  (i.e. {\sc clean()}) 
    These tasks were submitted as single work items, so avoided the known I/O bottleneck for accessing multiple datasets on \ple . The completion time for \aws\ for the \split\ operation was the worst for the three platforms and is dominated by the slow network reads for the EBS storage, whereas for the \clean\ operation the fast random access for the attached SSD disks give the best performance of the three platforms. \ple\ and \mgs\ have very similar performance for both tasks as they have similar disk I/O performance. 
    {The scatter, as derived from from the multiple passes through the processing, are indicated as ranges around the average values. They arise from the variations of workload and resource provisioning in the shared environments. Note the reduction of scatter on \aws\ between run time for \split\ and \clean\ due to the highly self-contained nature of the latter instance, and the increase in scatter for \mgs\ and \ple\ due to the greater random access requirements for  \split\ and \clean\ respectively.}
    }
    \label{fig:perf_time_comparison_img}
\end{figure}

We have compared the mean completion time of both Step 2 (\split) and Step 3 (\clean) on three computing environments as shown in Figure \ref{fig:perf_time_comparison_img}.

\subsection{\split}
The comparison of the three \split\ operations (on \aws, \mgs, and \ple) take as input the same visibility dataset - a single days observation of 400 GB, and produce as output a single 4MHz sub-band with an arbitrarily selected frequency range between 1020 and 1024 MHz. We repeated the same tests in three different days (over two weeks) in order to accommodate variations of workload and resource provisioning in the shared environments.

For a single instance of the \split\ operation, the \aws\ mean completion time (5346 seconds) is an order of magnitude longer than \mgs\ (437 seconds) and \ple\ (624 seconds). We hypothesised that this significant difference is attributed to the distinctive underlying I/O sub-systems in the three computing environments. Recall that both \ple\ and \mgs\ use Infiniband fabric (QDR) to interconnect compute nodes and storage nodes. In the case of \mgs, 70 GB/s read/write performance has been previously reported \citep{website:pawsey}. 
%
%
Since both input and output datasets are placed on these high performance storage nodes, the time spent on I/O is significantly reduced. On the contrary, the input for the \split\ operation on \aws\ is placed on the network-attached, general-purpose EBS disk storage volumes, which have no guaranteed I/O performance. To verify this conjecture, we used the system tool \strace\ to measure the I/O throughput on the \split\ operations.

Analysis of the \strace\ data
was from time series of I/O throughput (bandwidth) sampled every second for four types of I/O activities --- Sequential Write, Random Read, Random Write, and Sequential Read. 
Since the \split\ operation involves scanning the \msets\ by \casapy, sequential read/writes dominate these I/O activities. 
However, \mgs' peak sequential read throughput ($\sim$100 MB/s) for the \split\ operation is almost eight times greater than that on AWS EBS volumes ($\sim$12 MB/s). It is this difference of I/O performance that leads to the significant difference in the final completion time. The variance of the SPLIT completion time on \aws\ is significantly higher than \mgs\ and \ple. This is again caused by the network-attached EBS volumes, whose I/O performance is neither stable nor guaranteed in a multi-tenant network infrastructure. Likewise, the reason that the scatter of INVERT completion time on \mgs\ is far greater than \aws\ and \ple\ is attributed to the shared file system (i.e. the scratch space) relying on the resource-limited global storage network, where hundreds of users could concurrently perform I/O intensive workloads, significantly affecting one another.

\begin{figure}[h!bt]
    \centering
    \includegraphics[width=1.0\textwidth]{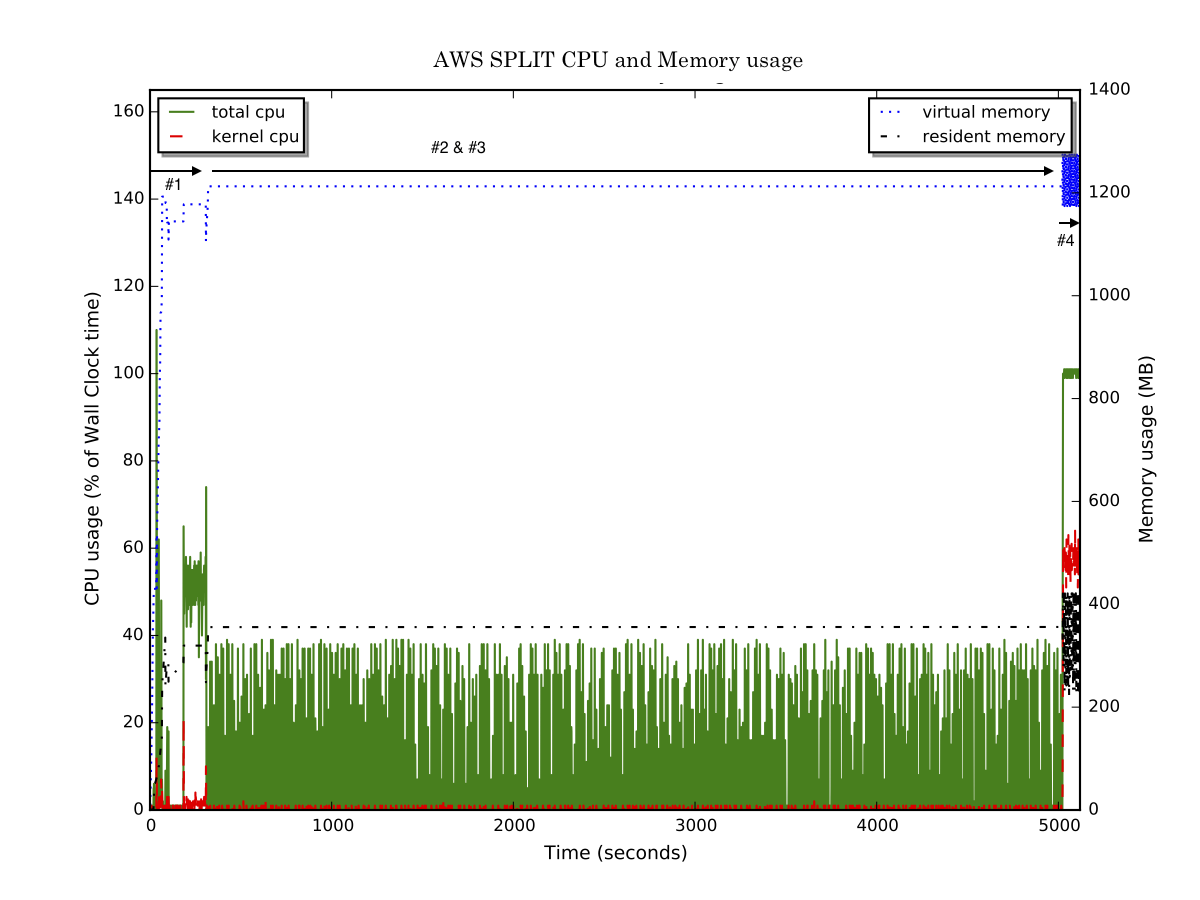}
    \caption{A plot of the cpu load (user and kernel, green and red solid line respectively) and memory usage (resident and virtual, dashed and dotted line respectively) as a function of time for the computing task. The breakdown into different stages of the task are indicated with the numbers and arrows. For the \split\ operation the relatively lower CPU usage (40\%) on \aws\  suggests that the CPU was idling, waiting for I/O requests in the queue to be completed}
    \label{fig:aws_split_cpu-mem_img}
\end{figure}
\begin{figure}[h!bt]
    \centering
    \includegraphics[width=1.0\textwidth]{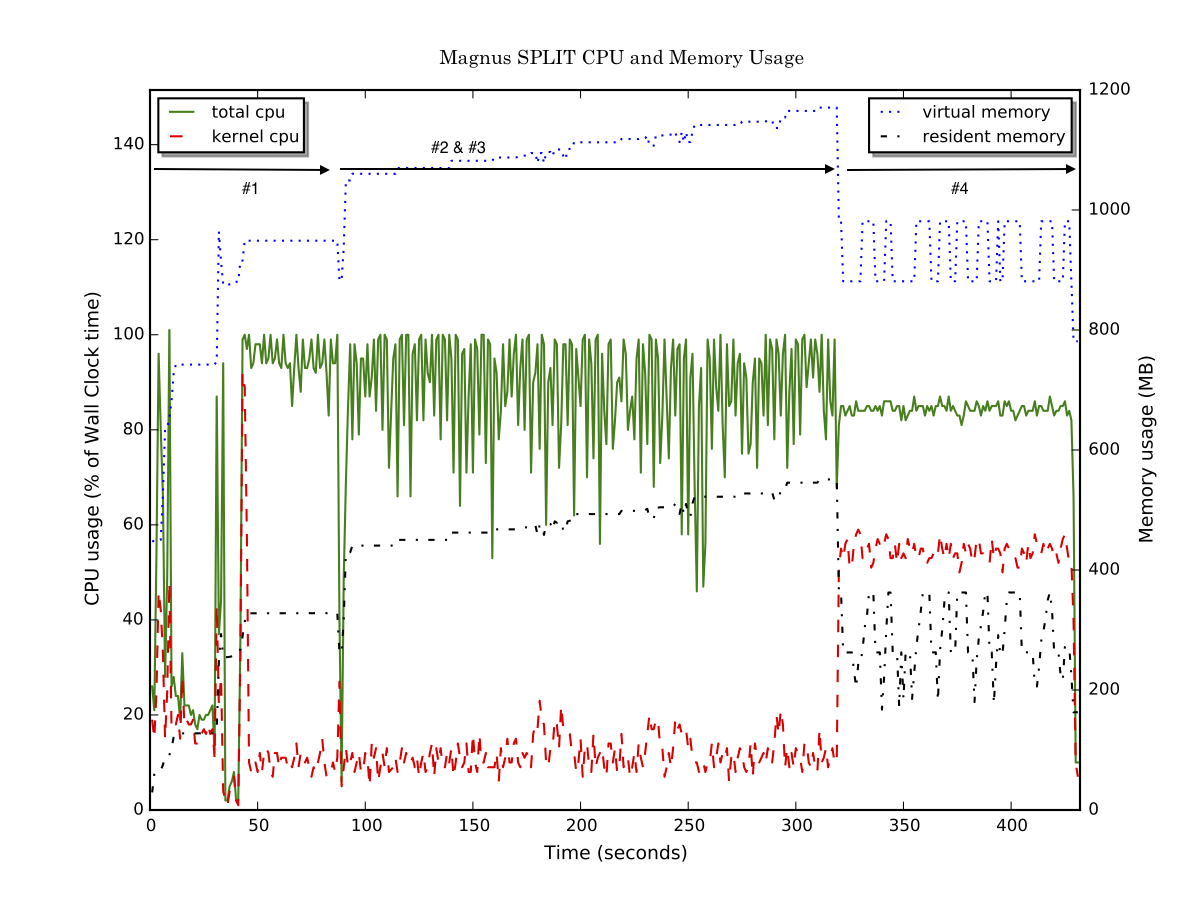}
    \caption{The relatively higher CPU usage (100\%) on \mgs\ for the \split\ operation suggests that CPU was busy, and I/O requests are dealt with fast enough to feed data to CPU. The labelling is as in Figure \ref{fig:aws_split_cpu-mem_img}.}
    \label{fig:magnus_split_cpu-mem_img}
\end{figure}
\begin{figure}[h!bt]
    \centering
    \includegraphics[width=0.9\textwidth]{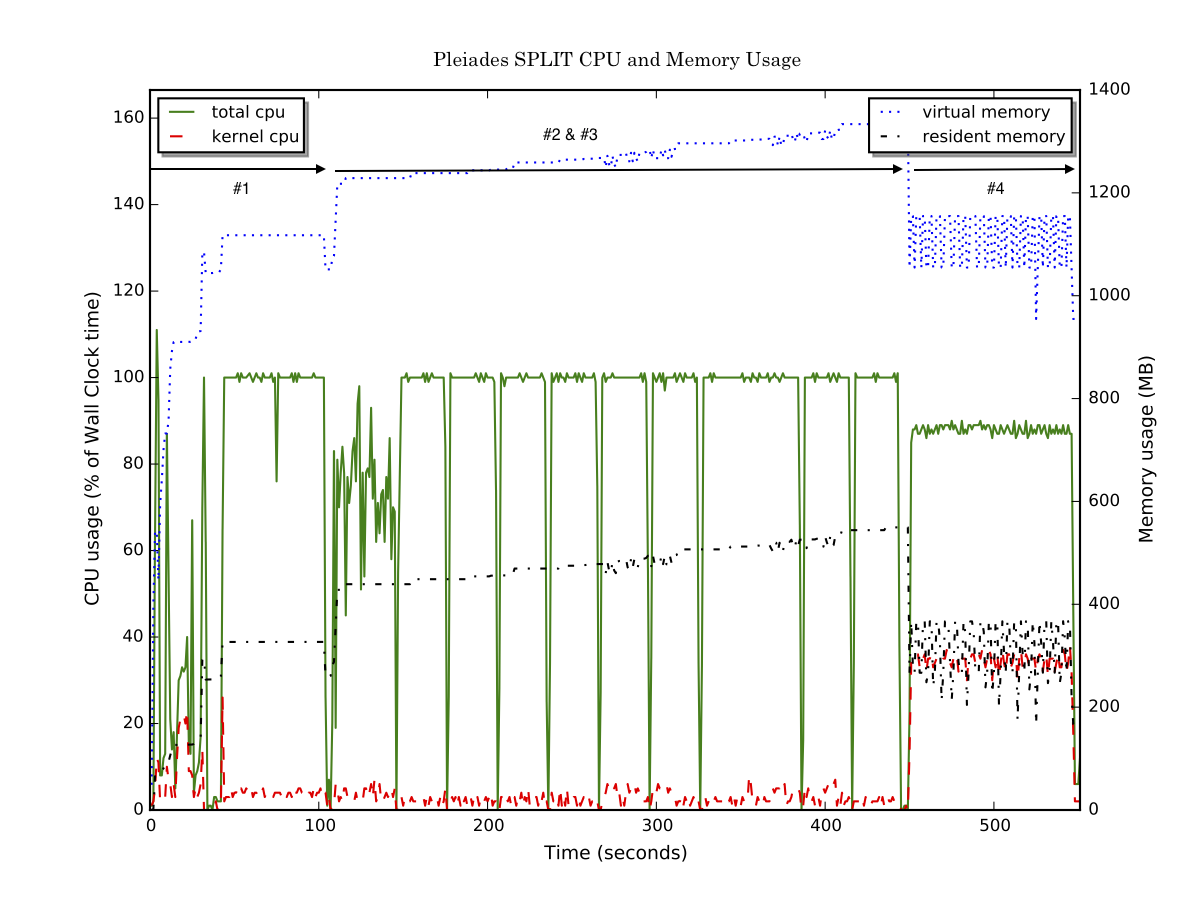}
    \caption{The relatively higher CPU usage (100\%) on \ple\ for the \split\ operation suggests that CPU was busy, and I/O requests are are dealt with fast enough to feed data to CPU most of the time. The labelling is as in Figure \ref{fig:aws_split_cpu-mem_img}.}
    \label{fig:pleiades_split_cpu-mem_img}
\end{figure}

To further examine the impact of I/O system on the completion time, we plot the CPU and memory usage of the \split\ operation on \aws, \mgs\ and \ple\ in Figure \ref{fig:aws_split_cpu-mem_img}, \ref{fig:magnus_split_cpu-mem_img}, and \ref{fig:pleiades_split_cpu-mem_img} respectively. Each figure shows CPU and memory usage for a particular run rather than the aggregated performance values across multiple runs over the two-week test period.

Compared to \mgs\ and \ple, the CPU usage on \aws\ is low --- less than 40\% (vs. 100\%), suggesting the CPU spends 60\% of its time waiting for I/O requests to be completed. 
Since the I/O is not fast enough to feed data to the CPU and memory, the actual (resident) memory usage on \aws\ is lower (less than 400 MB) than that on both \mgs\ and \ple, which reaches almost 600 MB at its peak. 
The underlying I/O system also caused noticeable differences between \mgs\ and \ple. 
As shown in Figure \ref{fig:pleiades_split_cpu-mem_img}  an extremely sharp ``trough" of CPU usage (as low as 0\%) appears frequently on \ple\ every 100 seconds or so.
However, CPU usage on \mgs\ (Figure \ref{fig:magnus_split_cpu-mem_img}) does not exhibit such spikes, only fluctuating between 40\% to 100\% after 90 seconds. This difference suggests that the \ple\ I/O system may have caused I/O waits that potentially created those sudden plunges on the \ple\ CPU usage curve. Consequently, a higher usage of CPU on \mgs\ has led to a slightly shorter overall completion time.

\subsection{\clean}
The three \clean\ operations take as input the same eight 4MHz sub-bands from 8 observations with an arbitrarily selected frequency range between 1136 and 1140MHz.  \aws\ (at 2967 seconds) performs better than both \mgs\ (6156 seconds) and \ple\ (5541 seconds) for two reasons. First, the \clean\ operation on \aws\ reads data directly from the ``local" SSD-backed ephemeral disks that are physically attached to the r3.4xlarge instance. 
Since reading data from directly-attached disks is much faster than from network-attached EBS volumes (as in the \split\ operation), the \clean\ performance has benefited greatly from the underlying I/O system. 
For single-node applications like our tests, \strace-based measurements 
show that the sequential I/O throughput of ephemeral disks is comparable to \mgs\ storage (the Lustre global file system) and slightly better than \ple\ storage (NFS V4-mounted, infiniband disk arrays). 
The random I/O performance on ephemeral disks is significantly better than \mgs\ and \ple\ since SSD disks are optimised for random access.


We found that the intensity of random I/O accesses in the \clean\ operation is much higher than that in the \split\ operation. 
This can be verified by the Inter-Arrival Time (IAT), which is defined as the time duration between the end of the previous I/O request and the start of the next I/O request. A smaller IAT corresponds to a more intensive series of I/O requests. IAT captures some intrinsic characteristics in an I/O workload regardless of its hosting platforms. 


In the \split\ operation the majority of random accesses have an IAT greater than 0.1 seconds, which is a thousand times longer than IATs of many sequential reads that reach \(10^{-4}\) seconds. On the contrary, small IATs in the \clean\ operations are primarily dominated by random reads and writes. 
The random accesses are almost three orders of magnitude more intensive in the \clean\ operation than in the \split\ operation. 
This explains why SSD-backed ephemeral disks perform better than HDD and network-backed storage on \mgs\ and \ple, leading to a shorter \clean\ completion time on \aws.

\begin{figure}[h!bt]
    \centering
    \includegraphics[width=1.0\textwidth]{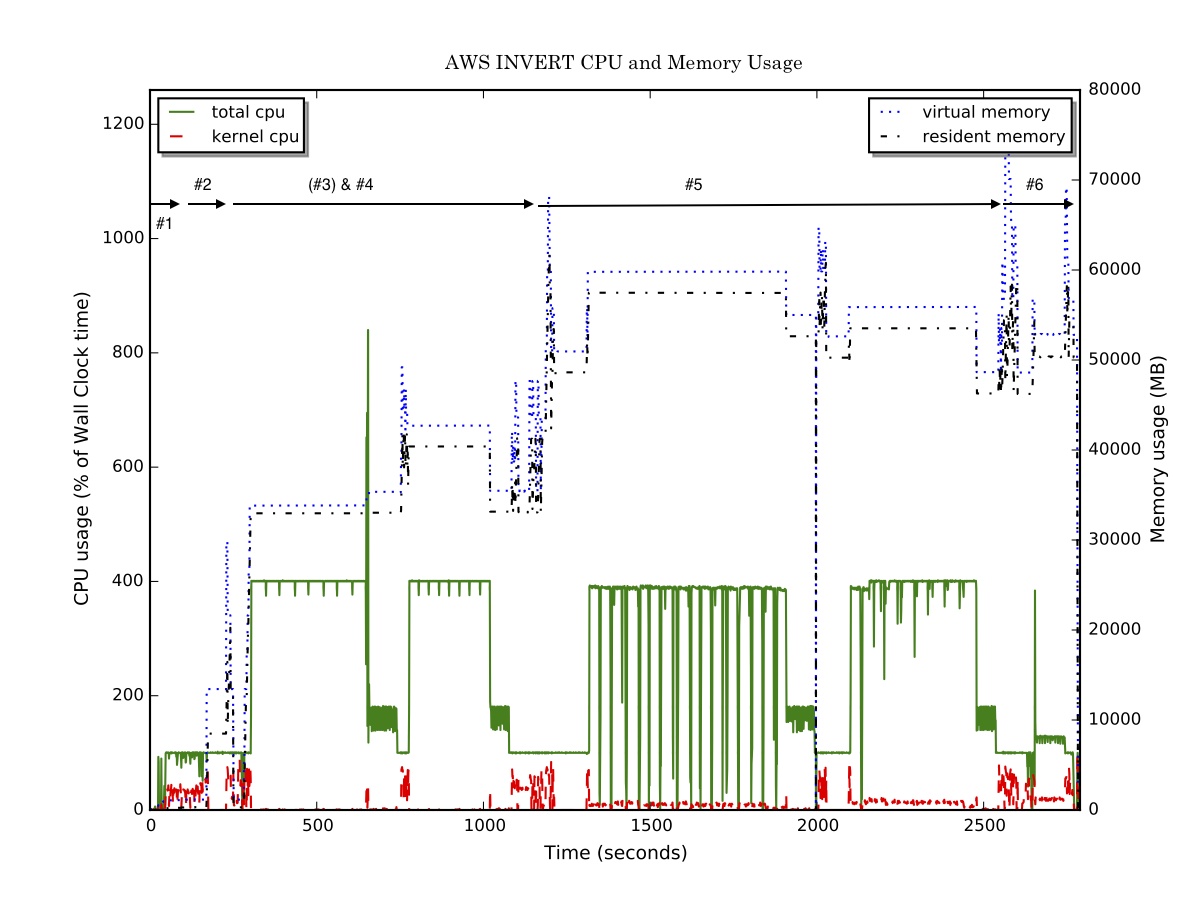}
    \caption{Four cores are utilised for the \clean\ operation on \aws, the memory consumption has peaked at 50GB. The labelling is as in Figure \ref{fig:aws_split_cpu-mem_img}.}
    \label{fig:aws_clean_cpu_img}
\end{figure}

\begin{figure}[h!bt]
    \centering
    \includegraphics[width=1.0\textwidth]{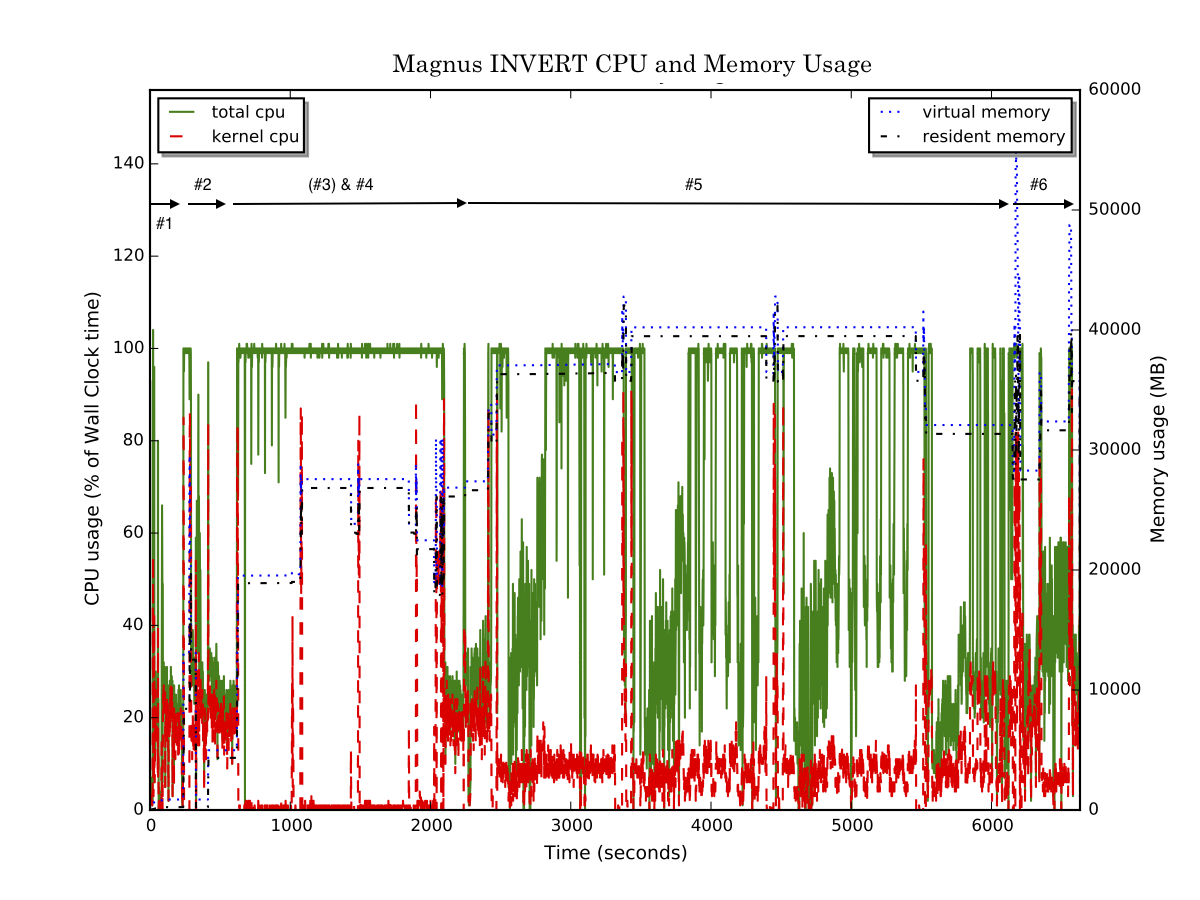}
    \caption{Only a single core is utilised for the \clean\ operation on \mgs. The labelling is as in Figure \ref{fig:aws_split_cpu-mem_img}.}
    \label{fig:magnus_clean_cpu_img}
\end{figure}

\begin{figure}[h!bt]
    \centering
    \includegraphics[width=1.0\textwidth]{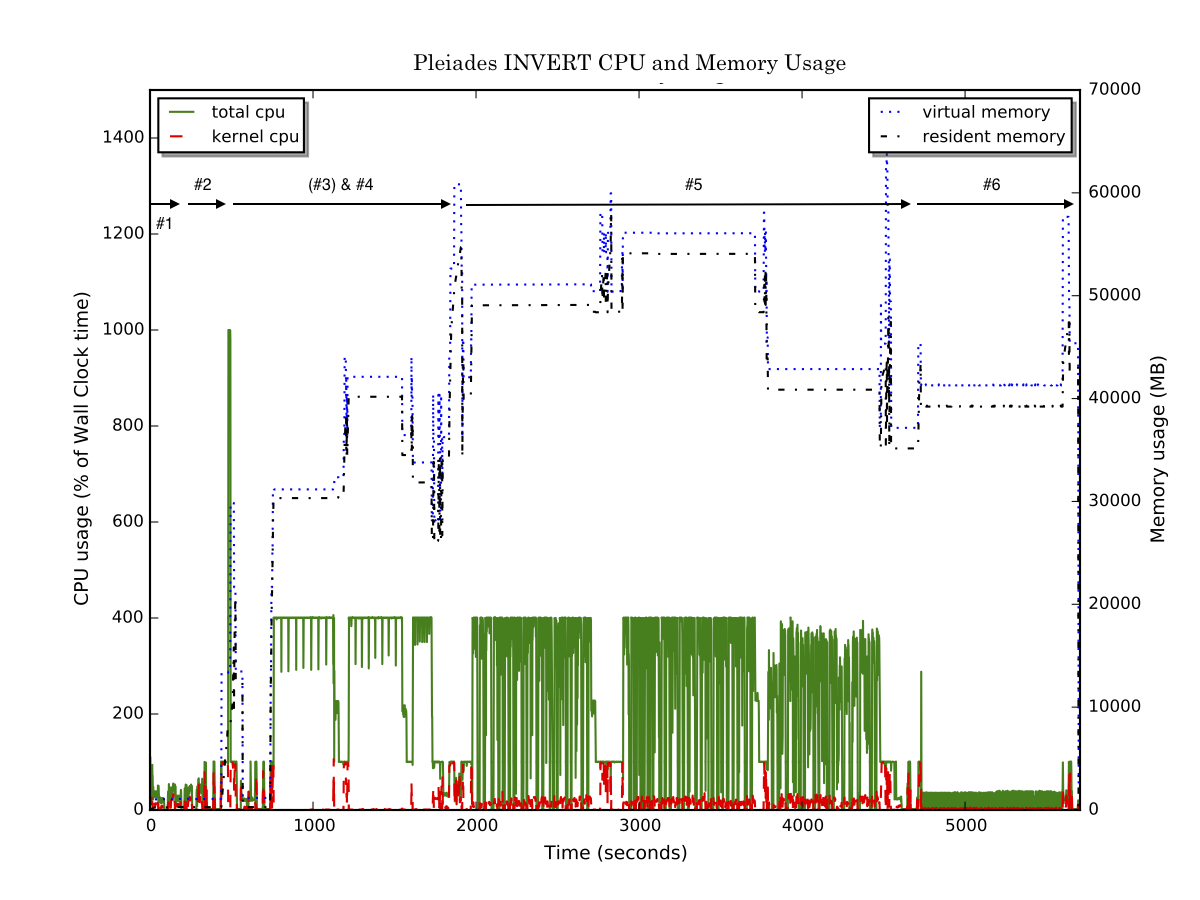}
    \caption{\clean\ on \ple\ can also easily exploit four cores, showing a 400\% CPU usage. The labelling is as in Figure \ref{fig:aws_split_cpu-mem_img}.}
    \label{fig:pleiades_clean_cpu_img}
\end{figure}

The second reason that \clean\ performs better on \aws\ than \mgs\ is due to the CPU usage --- four cores are fully utilized on \aws\ for the \clean\ operation (see Figure \ref{fig:aws_clean_cpu_img}), and we were only able to use a single core on each \mgs\ node (See Figure \ref{fig:magnus_clean_cpu_img}). This is due to the mismatch between the SLURM resource manager and the way \casapy\ program utilises multicores. Therefore it is not surprising to see that \clean\ on \aws\ is twice as fast as on \mgs. This also explains why \ple, which has a slightly weaker I/O system than \mgs, still performs better than \mgs\ as the \clean\ operation can fully utilise four cores on \ple, as shown in Figure \ref{fig:pleiades_clean_cpu_img}.


Note that for a ``fair" comparison, the \clean\ completion time on \aws\ also includes the data transfer time (287 seconds) between S3 and the ephemeral disks. This is because any practical use of ephemeral disks involves copying data from/to some non-volatile storage ``remote" to the \aws\ instances. 
Therefore, we need to account for data transfer to and from local storage for any useful computation. This is not the case for \ple\ and \mgs, where both input and output are persistent on global file systems accessible by other applications. Nevertheless the results show \aws\ still performs better, even after data transfer is accounted for. Furthermore, the S3 storage guarantees an availability of 99.999\%. Data on \mgs\ scratch space (without any fault-tolerance mechanisms) and \ple\ disks (RAID6) are subject to corruption and loss at a much higher rate.

Lastly, we note that \clean\ performance on \aws\ is much more stable than on \mgs\ (with a standard deviation of 877 seconds). We believe such stability arises from the directly-attached ephemeral SSD disks used exclusively by the r3.4xlarge \aws\ instance. This is in contrast to a shared global file system (e.g. that used in \mgs), where performance is susceptible to the constantly changing impact of jobs submitted by other users. In particular, in a workload dominated by random I/O accesses,
a large number of I/O operations have to travel across the storage network and get processed on a single Meta Data Server (i.e. the MDS in the Lustre filesystem), which can quickly become a bottleneck for processing I/O requests from thousands of running jobs\footnote{On a normal day, a ``squeue" command on \mgs\ shows more than 2000 jobs are either in the queue or running} at any particular point in time. Directly-attached SSD could also explain why \aws\ \clean\ performance is more stable than \aws\ \split, which uses network-attached EBS volumes subject to network traffic fluctuations within the AWS data centre. Moreover, the m3.xlarge virtual instance, on which \split\ runs, is more likely to share the same physical hardware resources with other virtual instances than is the r3.4xlarge instance, on which \clean\ runs. Therefore the \clean\ completion time is less likely to be affected by dynamically provisioned virtual machines.







\section{Validation}
\label{sec:resid}
\begin{figure}[h!bt]
    \centering
    \includegraphics[width=0.9\textwidth]{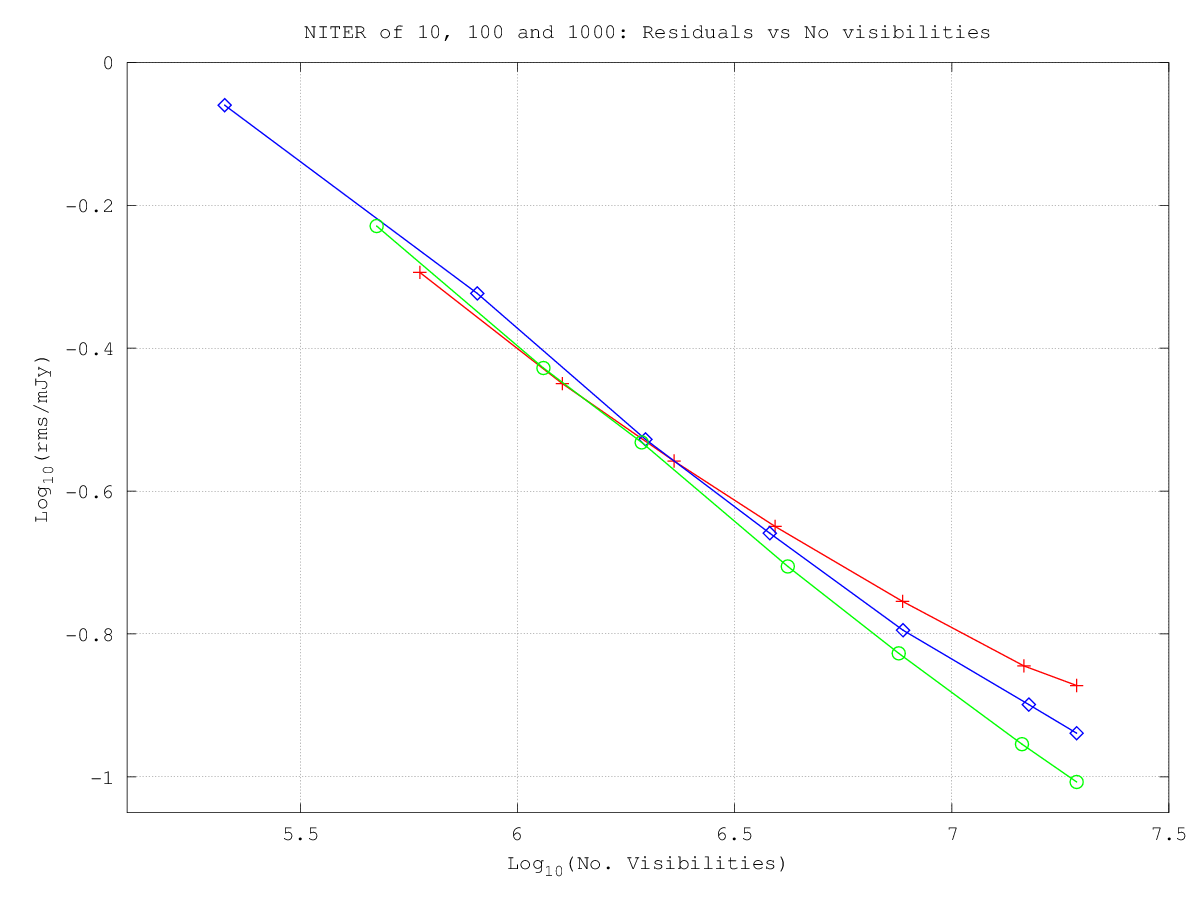}
    \caption{
      The log of the number of visibilities against the log of the residual RMS per channel (15.625kHz, but Hanning smoothed in the calibration pipeline) after including cleaning, with the number of iterations being 10, 100 and 1000, plotted in red with a cross, blue with a diamond and green with a circle respectively. The first point in each case is for a single days worth of data included in the imaging, the subsequent ones double until all the days are included.
      With a limited clean deconvolution of only 10 iterations the noise does not average down at the expected rate (rms $\propto$N$^{-0.5}$), only achieving a factor of -0.38; when more deconvolution is included this is significantly improved, reaching -0.45 and -0.49 for 100 and 1000 iterations, respectively.}
    \label{fig:noise}
\end{figure}

To confirm that the pipeline was functioning correctly we plot the rms of the image residuals in Figure \ref{fig:noise} as a function of the number of visibilities (with 8 second integrations) included in the inversion. 
For a Gaussian distribution of noise we expect the rms of the residuals to decrease with the square root of the number of visibilities. In these tests we firstly ran the \clean\ step with 10 iterations of deconvolution included. We then calculated the rms of the residual image (i.e. the source subtracted, nominally noise-only cube). We performed this measurement on the residual image for one days worth of data and one sub-band and then repeated it, doubling the number of days thereafter. 
The slope in this case is -0.38 and is noticeably non-linear, which indicates that the residuals are non-Gaussian. This we hypothesised was from the residual sources uncovered as the cube reaches a higher sensitivity.
Therefore we repeated the analysis with the number of clean iterations set to 100 and then 1000. For these cases the slope was -0.45 and -0.49, respectively. This gives us confidence that firstly the pipeline is operating correctly and also that deconvolution will be required in the final workflow. The noise level achieved is in line with 
the expected sensitivity, based on the VLA calculator.

Additionally we checked the completion time for the \clean\ stage against the number of datasets. We find that it scales very close to the square root of the number of visibilities, underlining the advantage of imaging all days together rather than each day individually and summing these for the final data product. On the other hand we find that the completion time as a function of the number of channels imaged is linear, so doubling the number of channels doubles the completion time. 

\section{Summary of metrics and operations}

In the following sections we summarise the contributions to the total time taken to complete the different operations on the different environments. 









\subsection{Inter-arrival times metric}

The Inter-Arrival Time (IAT) is the time between I/O requests. The lowest values therefore represent the most intensive I/O workloads. In the majority of cases the I/O request is only issued once the preceding work-task is completed, therefore it also represents the non-I/O limited computational speed. For \split\ the sequential reads are the dominant load, arriving on all machines every 0.1~mseconds. \split\ has a low CPU intensity, and deals with the data sequentially, so this behaviour is as expected. In comparison the \clean\ operation places the highest demand on random reads, which is why computing environments with SSD disks (i.e. our selected \aws\ configuration in the \clean, but not in the \split, operations) perform so much better than systems with global file-systems.

\subsection{I/O Throughput metric}

When massive datasets are being accessed, as for this project, the task completion times become extremely sensitive to I/O requirements of the various sub-tasks. We find the throughput has very similar behaviour to that of the IAT; \split\ is  dominated by sequential reads, and the infiniband backed global filesystems work very well. On \mgs\ and \ple\ the read maximum throughput ($\sim$100MB/s) approached the theoretical performance. On \aws, however, the maximum throughput was almost an order of magnitude less. This is a consequence of running on \aws\ network disk instances. To have a large ephemeral disk on \aws\ is expensive, therefore we chose not to use this option for the \split\ stage.
The last stage of the \split\ on \aws\ involves the transfer of the flag tables, which involves a massive read of the whole table for a minuscule write of the relevant section. This could easily be improved by dropping flagged data at ingest. 
Generally the writing of output is less important than the reading, as much more data is read in than written out. This is discussed in Section \ref{sec:drt} and could be improved in future \split\ operations, which could potentially make the write speeds an issue. 
The \clean\ operation, for which random reads as significant, performed poorly on the Infiniband based disk solutions but extremely well on the \aws\ SSD-backed processing. In this case we did require a high powered compute platform so it made sense to use one with a large SSD scratch space (i.e. the r3 instance). 

\subsection{CPU load metric}

\casapy\ does not support massively parallel operations, although some tasks support OpenMP for up to four cores. However we failed to get this working on the \mgs\ system because of difficulties parsing environment variables under the SLURM queue manager. Therefore we found that it was impossible to achieve the CPU-loads on \mgs\ for \clean\ that we could achieve on \ple\ and \aws. For the \split\ operation the computational load is single threaded, so the maximum load would be 100\%, but this is potentially throttled by the I/O throughput. For the \aws\ the limited network bandwidth to the EBS disks prevents us achieving more than 40\% CPU load. In comparison the CPU load on \mgs\ and \ple\ reaches 100\% as the sequential reads over the Infiniband are well provisioned.

\subsection{{\sc mstransform()} Task}

The {\sc mstranform()} task has four separate stages, all of which have significantly different access and usage patterns. These stages are:
\begin{enumerate}
\item Selection: The requested data (one or more spectral windows) is selected. 
\item Regrid: The requested data is FFT-ed, shifted and inverse FFT-ed. 
\item Apply: The data channels required are selected from the regridded data.
\item Flagging: Any flagging updates required are applied
\end{enumerate}

1) The selection step (typically about 100 seconds) reads the input file information and prepares the new MS for output. Processing is dominated by sequential reads.
2) The regrid and 3) apply processes are CPU bound, unless the sequential reads for new data are provisioned slowly, as is the case on \aws.
4) The final $\sim$100 seconds updates the new MS flagging table from the old one. This is a read only process in our investigation as there were no flags to be transferred. 

\subsection{{\sc clean()} Task}

The {\sc clean()} task has upto five separate stages, all of which have significantly different access and usage patterns. These stages are:
\begin{enumerate}
\item Creation: The requested image is created and prepared.
\item Gridding: The channels to be imaged are gridded
\item Major Cycle: If there are clean iterations requested there will be a loop where the models are converted to the visibility sampling (degridding), subtracted from the original visibilities, after which the residuals are gridded again for a further cycle of clean.
\item Deconvolve: The image plane is iteratively deconvolved with a peak search and subtraction.
\item Restore: The image plane is restored by replacing the removed model components with a Gaussian beam convolved with those model components.
\item Finalise: The image is written out.
\end{enumerate}

1) The creation step (typically about 70 seconds) prepares the new image file for writing and clears the model fields. Processing is dominated by sequential reads and writes. 
After this the memory is flushed. 
2) The antenna response is computed and the required gridding array allocated. This involves building the dirty image, whether or not any clean iterations are performed. This stage is normally CPU limited with moderate sequential and low random read requirements. 3) If clean iterations are requested additional cycles of processing occur in this section, as the model components are stored in memory. 
4) Next the deconvolution and 5) restore steps are performed, which occurs even if no clean iterations are requested. This step is characterised by both sequential reads and random writes; on the machines with infini-band provisioned disk the writes are several fold slower than the reads, unlike on the SSD-provisioned \aws\ machine. This accounts for the major performance gain of \aws\ over the other systems. 
6) The final stage is to write out the results, and is characterised by a high sequential write and low CPU demand. 

The validation shows that i) the time for completion is linear with the number of channels and follows the square root of the number of visibilities and ii) to achieve the theoretical noise we will require deconvolution in the imaging stage. The former point led us to image all days together but divide the data into manageable sized sub-bands, the latter point will influence the final workflow design.

\section{Considerations for Workflow Design}

We have successfully explored the options for the computing platform for the \chiles\ imaging pipeline, and constructed a workflow solution. 
The exact details of this implementation are hard to convert into definitive long-living prescriptions for success for other projects because of: the short term nature of computing infrastructure, the different access patterns for other computing tasks and even the improvements which will be made to our own computing tasks based on the identification of the processing bottle necks in these investigations.
%
{Nevertheless we believe the result and methods presented could be useful for other investigators to configure their in-house computing/storage environment or to formulate cost-effective Cloud strategies suited to workloads similar to the \chiles\ imaging pipeline described in this paper. Therefore, we provide in Table \ref{tab:perf_summary} a summary of Section 6 and Figures 5 --- 10.}
%
The considerations will include the resources available to the user, the scale of the compute required and the difficulty of transferring the data to and from the compute environment. 
Therefore we have ranked the following considerations and performance measures on a scale from 0 to 5 (unacceptable, poor, passable, acceptable, good, excellent). For costs, high costs are considered poor and lower costs would be scaled higher.

\subsection{Costs}

The total cost of ownership in our three environments are very different. 
For a fully-owned cluster the purchase has to be made ahead of time, and then the system needs to be maintained. 
The ICRAR cluster, \ple, was purchased 4 years ago at a cost of $\sim$AUD\$50K for the six compute nodes and we have one full time staff who is responsible for the management. 
It supports all the users in ICRAR, but not beyond. However, as it could not complete the pipeline in parallel (which therefore means the system could not perform the required data reduction) we cannot measure hours of operation required and therefore cannot estimate its operational cost. If the tasks were perform sequentially we estimate it would take 1,060 hours.

\mgs\ was purchased one year ago at a cost of $\sim$AUD\$12M for the 1536 compute nodes and 3PB of disk storage. For total cost of ownership, the Pawsey Supercomputing Centre uses a figure of AUD\$0.67 per node hour for compute jobs~\citep{email:GeorgeBeckett-2015-04-30}.
It supports many users who apply for time in regular calls for proposals; our tasks would request 44 nodes, $\sim$3\% of the total capacity, so we use 3\% of the total costs for comparison.

The \aws\ system, on the other hand has no setup costs for the user, but computational usage is billed monthly. 
For the \chiles\ data reduction of the first epoch including debugging, the total AWS bill for computation and storage was $\sim$AUD\$2K. This breaks down into \$225 for $\sim$3,000 hours of computation, \$1.2K for 7TB of on-demand EBS storage and \$40 for 65TB of long term S3 storage. This demonstrates the potential for low cost computing provided by cloud facilities, but also the potential for cost blow out if large amounts of data are kept long term in high availability storage. 
We have ranked the three systems (\aws, \mgs\ and \ple) for capital costs as `excellent', `passable' and `good' respectively and for operational costs as `excellent', `excellent' and `unacceptable' respectively.

\subsection{Usability}

The issue of `Usability' and `Control' in some fraction reflects on our limitations as computer users, rather than the intrinsic capabilities of the machines. 
This we break down into two branches: the amount of control we had over the system and the ease of use of the system. 
The importance of control is demonstrated by the significant hit in performance we had on the Pawsey machines, because we could not reconfigure nor test the systems.  Without root access, which is the case with \mgs, we could not fully implement the performance measurements or resolve the environment variable problems, for example. On both our own system \ple\ and on the \aws\ cloud we had full root access, which allowed us to maximize the performance. 
On the other hand the setup and submission of the jobs on \mgs\ and \ple\ were much simpler compared to those on \aws, where we have to search for the best moment to launch instances, to configure those instances on the fly and to access data products spread over a complex zoo of support infrastructure. We have ranked the three systems (\aws, \mgs\ and \ple) for control as `excellent', `acceptable' and `excellent' respectively and for usability as `passable', `good' and `good' respectively.


\subsection{Data Transfer}
It is a major overhead to transfer the massive \chiles\ datasets from NRAO onto the local clusters and the supercomputing centre in ICRAR, Pawsey and \aws\ in Sydney, where it can be processed by the compute platform. Both \ple\ and \mgs\ have physical 10Gb network interfaces to the outside world, which provides reasonably good data transfer infrastructure. \mgs\ has set up a data transfer service on two dedicated data nodes with optimal configuration for both inbound and outbound traffic. The \ple\ cluster also has a dedicated data transfer node with a 10 Gb NIC, but with only two CPUs and 8 GB of RAM, which means a relatively smaller buffer capacity during data transfer compared to \mgs. Nevertheless, we found that data transferring time between ICRAR and Pawsey to be acceptable as long as we managed to saturate the link. However, data throughput recorded on our \aws\ copy machine has only reached 1Gb for parallel streams during data transfer. Moreover, it is unknown (i) whether this link/NIC is exclusively used by our application and (ii) whether the bandwidth has been deliberately ``shaped'' by AWS, since the throughput also fluctuates considerably when we increase the number of parallel stream from 1 to 2, then to 4 as shown in Figure \ref{fig:aws_network_thruput_img}. Compared to \mgs, an advantage of \ple\ and \aws\ is the root access, which allows us to fine tune the Linux kernel (e.g. increase the default TCP window buffer) for optimal data transfer. Given the above analysis, we have ranked the three systems (\aws, \mgs\ and \ple) for data transfer as `Acceptable', `Good' and `Good' respectively.

\subsection{I/O Performance}
For workloads (such as \split ) dominated by sequential I/O read/writes, the EBS volumes used in \aws\ performed almost ten times worse than both \mgs\ and \ple\ in terms of throughput during the processing. The same is true for IOPS achieved by \split\ on (EBS-backed) \aws\, which is again an order of magnitude smaller than those on \mgs\ and \ple . However for random I/O intensive workloads such as \clean, the SSD storage used in \aws\ instances has definitely shown the best throughput results (up to 500 MB/s). Similarly, the IOPS for \aws\ \clean\ has peaked at \(10^4\), five or ten times higher than the other two. Overall, we have ranked the three systems based on peak performance. For the bandwidth we ranked \aws, \mgs\ and \ple\ as `Good', `Acceptable' and `Acceptable' respectively and for I/O performance as `Excellent', `Good' and `Good' respectively.

\subsection{Radar Analysis}

\begin{figure}[h!bt]
\centering
\includegraphics[width=0.8\textwidth]{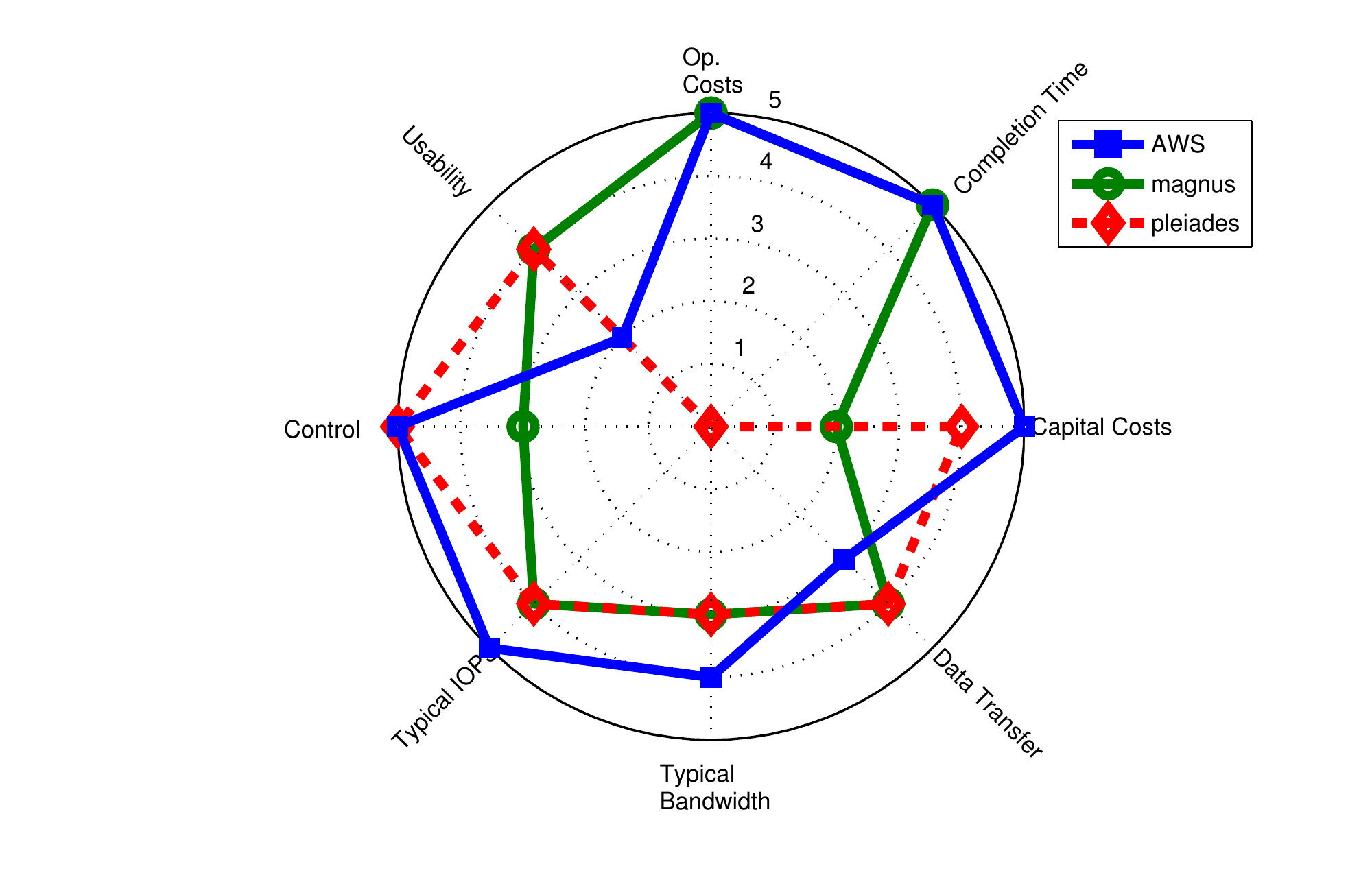}
\caption{Spider plot for the workflow on the three platforms under test, \aws\ (blue line with squares), \mgs\ (green line with circles) and \ple\ (red dashed line with diamonds) respectively. The ranking in Table \ref{tab:spider}, from 5 to 0, is used along the corresponding labelled axes.}
\label{fig:spider}
\end{figure}

Table \ref{tab:spider} summarizes these performance considerations for the platforms and plots them on a radar plot in Figure \ref{fig:spider}. 

\paragraph{Moderate Size Departmental Computing} \ple, red dashed line with diamonds. 
The cluster was unable to perform the parallel pipeline analysis, as highly parallel tasks fail because of the I/O limits. Notwithstanding it was essential for testing the work-units. 

\paragraph{High Performance Computing} \mgs, green line with circles. 
Not surprisingly the Cray XC40 was the fastest and the per-node compute costs are very reasonable. Nevertheless it is an inflexible environment, because it is a shared national resource and is a monolithic architecture. We can not adjust the computing to the problem and we have no root access to tune the performance to our needs. These are natural consequences of using such a shared resource. 

\paragraph{Cloud Computing} \aws, blue line with squares.
The ability to tune the hardware to the particular problem was the strongest advantage of the \aws\ system. 
For example the \ple\ data reduction bottle neck was clearly the I/O to the single disk when running parallel tasks.
This was initially also true in the \aws\ implementation, but by upgrading the instance for that work-unit (to provisioned SSD disks with high IOPS) we were able to easily improve the performance. 
The range of available hardware options for the implementation of different aspects of the workflow is one standout advantage of cloud computing approaches.


\begin{table}[h]
\centering
\hspace*{-0.5cm}\begin{tabular}{r||c|c||c|c||c|c}
\hline
Consideration & \multicolumn{2}{|c||}{\aws}  &  \multicolumn{2}{|c||}{\mgs} & \multicolumn{2}{|c}{\ple} \\
 \hline
Completion Time &  96hr & 5 & 110hr & 5 & 1,060 hr (est.) & 0\\
Capital Costs &  \$0 & 5 & \$340,000 & 2 & \$50,000 & 4\\
Operational Costs &  \$2,000 & 5 & \$3,240 & 5 & - & 0 \\
Data Transfer & 1Gb (high variance) & 3 & 10Gb & 4 & 10Gb & 4 \\
Typical Bandwidth & $\sim$300MB/s & 4 & $\sim$100MB/s & 3 & $\sim$100MB/s & 3 \\
Typical IOPS & $\sim$1,000 & 5 & $\sim$100 & 4 & $\sim$100 & 4 \\
Control & Root Access & 5 & Limited Access & 3 & Root Access & 5 \\
Usability & Python/Boto & 2 & Python & 4 & Python & 4 \\
\hline
Product ($\Pi/5^8$) && 0.15 && 0.07 && 0 \\
\end{tabular}
\caption{The performance rankings for the workflow items on the three platforms under test, \aws, \mgs\ and \ple\ respectively. The metric is given for each aspect, and is ranked, from 5 to 0, as `Excellent', `Good', `Acceptable', `Passable', `Poor' or `Unacceptable'. 
}
\label{tab:spider}
\end{table}

\begin{table}[h]
\centering
\hspace*{-0.5cm}\begin{tabular}{l||llll||l}
\hline
Operation& Platform & Peak Memory & I/O Throughput & CPU Usage & I/O Characteristics \\
\hline
       & \aws\ (EBS) & 420 MB & $<$10MB/s & 40\% & Sequential \\
       \cline{2-5}
\split & \mgs & 545 MB & 40 $\sim$ 100 MB/s & 100\% & read/write \\
       \cline{2-5}
       & \ple & 390 MB & 60 $\sim$ 100 MB/s & 100\% & dominated \\
\hline
       & \aws\ (SSD) & 60 GB & 70 $\sim$ 500MB/s & 400\% & Random writes \\
       \cline{2-5}
\clean & \mgs & 30 GB & 50 $\sim$ 400 MB/s & 100\% & and sequential \\
       \cline{2-5}
       & \ple & 35 GB & 50 $\sim$ 300 MB/s & 400\% & reads dominated \\
\hline
\end{tabular}
\caption{Performance summary broken down for the \split\ and \clean\ operations across three measured metrics --- peak memory usage, I/O throughput, and CPU usage. Inherent I/O characteristics for each operation are also summarised in the last column. The profiling information in this summary constitutes an essential input for optimal resource provisioning and job scheduling.}
\label{tab:perf_summary}
\end{table}

\subsection{Future Developments}
We are using these studies to refine and develop our operating infrastructure. We list here the improvements we are making for processing the second epoch of \chiles\ data, as informed by the performance measurements made. These are improvements which are probably of interest to all facility managers. 
\begin{itemize}
\item SSD for local high speed scratch space. We are installing local SSD disks on all nodes of \ple\ as that will allow a high-speed random access on locally-hosted data files. With this we maybe able to complete the processing on a moderate sized cluster.
\item Improved I/O performance. Conversations with AWS are underway to improve the I/O limitations we have been struggling with.
\item Trialling the Next Generation Archive System {\sc ngas} \citep{ngas, ngas_mwa} for the transfer of the data from NRAO to the \aws\ infrastructure. We will attempt to perform the entire data reduction chain, including flagging and calibration, on the cloud computing platform.
\item Developing a data-driven workflow for the \chiles\ project, which will be able to prototype many of the SDP concepts and pipeline designs. 
\item A new task is being developed in \casapy , {\sc uvgridder()} can cumulatively grid all days onto one uv-grid, which may prove to be the best approach \citep{uvgridder,dingo_memo}.
\end{itemize}

\section{Acknowledgements}

The Karl G. Jansky Very Large Array and the National Radio Astronomy Observatory is a facility of the National Science Foundation operated under cooperative agreement with Associated Universities Inc.. We wish to thank the CHILES team for flagging and calibrating the data used in these tests. 
This work was supported by grants from Amazon Web Services, the AstroCompute project and by resources provided by the Pawsey Supercomputing Centre with funding from the Australian Government and the Government of Western Australia. The \mgs\ supercomputer \citep{website:pawsey} is managed by The Pawsey Supercomputing Centre. The CHILES project is supported by a collaborative research grant from NSF.

\end{document}